# A Deployment-Oriented and Resource-Efficient Hybrid Neural-Symbolic Framework for Explainable DDoS Detection in Operational Technology Environments

Mikiyas Alemayehu, *Student Member, IEEE,* Mohamed Chahine Ghanem, *Member, IEEE,* Kevin Curran, *Senior Member, IEEE,* Hamza Kheddar, *Senior Member, IEEE,* Ioanna Dionysiou, *Senior Member, IEEE,* Elhadj Benkhelifa, *Senior Member, IEEE,* and J. J. Garcia-Luna-Aceves, *Fellow, IEEE*

***Abstract*—This paper presents a hybrid neural-symbolic detector of distributed denial-of-service (DDoS) attacks on the availability of operational technology (OT) networks, designed for their resource-constrained gateways, and an evaluation matched to deployment. A compact gated recurrent unit (GRU) is fused at score level with a shallow, readable decision tree, and weight and threshold are selected on validation data under an explicit false-positive budget, whose attainment in finite samples is analysed. The evaluation on CIC-DDoS2019, Edge-IIoTset and CICIoT23 uses contiguous or capture-held-out partitions, dependence-aware intervals, and precision and alert load at operational prevalence. Over ten seeds at a 1% budget, the hybrid's false-negative rate on CIC-DDoS2019 is 0.08%, against 0.26% for the GRU and 0.17% for the tree. With a day of mostly unseen attack families held out, the tree alone carries the decision. Audits bound these claims: Edge-IIoTset is saturated by duplicates, no benchmark as used provides interleaved traffic in capture order, and an unconstrained tree is more accurate on two of them, so the case for the hybrid rests on short, stable rules and a measured footprint. On a Siemens SIMATIC IoT2050 gateway, tested with Modbus/TCP and MQTT floods, the INT8 model needs 2.9 ms per window, 31 MB of memory and 3.5 mJ per inference.**

## NOMENCLATURE

| | |
|---|---|
| CART | Classification and regression tree. |
| FPR, FNR, TPR | False-positive, false-negative and true-positive rates. |
| MBB, CB | Moving-block and cluster bootstraps. |
| $L$, $s$ | Window length and stride, in records. |
| $d$, $h$ | Input dimension and hidden units of the GRU. |
| $p_n$, $p_s$, $p_h$ | Neural, symbolic and hybrid scores. |
| $\alpha$, $\tau$ | Fusion weight and decision threshold. |
| $\beta$, $\delta$ | False-positive budget; probability of exceeding it. |
| $n_0$, $k$ | Benign validation windows; $k = \lfloor \beta n_0 \rfloor$. |
| $n_{\mathrm{eff}}$ | Effective sample size. |
| $\pi$ | Attack prevalence. |
| $\lambda$, $r_{\mathrm{dec}}$ | Flow rate and decision rate, per second. |
| $T_{\mathrm{exp}}$, $T_{\mathrm{fill}}$, $T_{\mathrm{inf}}$ | Export, window-fill and inference times. |
| $T_{\mathrm{det}}$, $T_{\mathrm{max}}$ | Detection delay; budget for $T_{\mathrm{inf}}$. |

Manuscript received XXXX; revised XXXX. This research received no external funding. *(Corresponding author: Mohamed Chahine Ghanem.)*

M. Alemayehu is with the School of Computer Science and Mathematics, Keele University, Keele ST5 5BG, U.K.

M. C. Ghanem is with the CSchool of Computer Science and Mathematics, Keele University, Keele ST5 5BG, U.K., and also with the Cybersecurity Institute, University of Liverpool, Liverpool L69 3BX, U.K. (e-mail: m.ghanem@keele.ac.uk).

K. Curran is with School of Computing, Engineering and Intelligent Systems, Ulster University, Londonderry, U.K.

H. Kheddar is with the LSEA Laboratory, Department of Electrical Engineering, University of Medea, Medea 26000, Algeria.

I. Dionysiou is with the Department of Computer Science, University of Nicosia, Nicosia, Cyprus.

E. Benkhelifa is with the CSchool of Computer Science and Mathematics, University of Toronto, Liverpool John Moores University, Liverpool, L3 5AH, U.K.

J. J. Garcia-Luna-Aceves is with the Centre of Excellence for Networking Innovation in Toronto, University of Toronto, Toronto, ON M5S 3G4, Canada.

## I. INTRODUCTION

OPERATIONAL technology (OT) networks connect programmable logic controllers (PLCs), supervisory control and data acquisition (SCADA) servers and field devices in power, water, process and manufacturing plants. Their convergence with IP networks exposes them to distributed denial-of-service (DDoS) attacks that saturate links and delay or drop control and telemetry traffic [1]–[4]. Such an attack is aimed at availability, which IEC 62443-3-3 lists among its foundational requirements for industrial systems [5]. The natural vantage point for detecting such attacks is the boundary between the plant and the enterprise network, and the computers available there are industrial gateways: fanless Arm devices with one or two gigabytes of memory, no accelerator and a power budget of a few watts. Deep detectors reach high accuracy on public benchmarks but are demanding on hardware of this class [6], and even edge-optimised variants [7] give an operator no reason for an alert. Rule-based detectors are cheap and readable, yet they miss patterns that appear only across several flows.

Neuro-symbolic intrusion detection, which couples learned representations with symbolic reasoning, has been proposed as a way out of this trade-off [8]. Most published designs place differentiable logic, knowledge graphs or solvers in the inference path and target general-purpose networks. For deployment in OT they leave three gaps.

*1. Evaluation:* Record-level shuffled splits scatter adjacent flows of one capture across training and test data. For a detector that reads windows of consecutive flows this is a form of data snooping [9]–[11], and it leaves the windows without temporal meaning. Operating points are usually chosen to maximise F1 on attack-dominated test sets, which says little about precision when attacks are rare [12].

*2. Interpretability:* Explanations are illustrated with example rules. The rules that a model has actually learned are seldom listed in physical units, compared with the attack mechanism, or tested for stability under retraining.

*3. Deployment:* Latency, memory and energy are rarely measured on industrial hardware, and model inference time is often presented as if it were the time to detect an attack.

The three gaps have a common root. A detector at the plant boundary is a component of the system it monitors, and its error rates are reliability quantities: a missed flood leaves the plant exposed, and false alerts wear down the operator's trust until alerts are ignored. Such rates have to be specified, verified on data that resemble operation and reported with their uncertainty. They should not be a by-product of maximising accuracy on a shuffled benchmark.

This paper addresses the three gaps with a deliberately simple detector and a deliberately strict evaluation. The detector fuses the score of a two-layer gated recurrent unit (GRU) over a window of 16 flow records with the score of a decision tree of depth at most five applied to the latest record. The branches are trained independently and combined by a convex weight; weight and threshold are selected on validation data by a Neyman-Pearson rule that maximises recall under an explicit false-positive budget. The tree makes every alert traceable to a short conjunction of threshold tests. The design is a loosely coupled hybrid, and we claim neither differentiable logic nor joint training. The contributions are as follows.

- *Detector and operating point:* A score-level fusion of a compact GRU and a shallow tree whose operating point is selected under a false-positive budget by one rule applied identically to every model compared, with the finite-sample behaviour of that rule and a closed-form parameter and operation count (Section IV).
- *Leakage-controlled protocol and benchmark audits:* The positive class is restricted to DoS families, windows never cross a partition boundary, partitions are contiguous blocks with a purge gap or whole held-out captures, and all fitting uses training data only. Audits of record order, identifiers, feature concentration and duplicates establish what each benchmark can support (Section V).
- *Dependence-aware statistics and operational reading:* Block and cluster bootstrap intervals with an effective sample size replace binomial intervals on overlapping windows, and operating points are translated into precision and false alerts per hour at operational prevalence (Sections V-D and V-E).
- *Evidence for interpretability:* The learned rules are reported in physical units with support, purity, stability over retraining, and counts of the windows on which fusion corrects or breaks the neural decision (Section VI-F).
- *Measurements on an industrial gateway:* An INT8 deployment is profiled for latency, memory and energy on a Siemens SIMATIC IoT2050 and a Raspberry Pi 5 and exercised with six Modbus/TCP and MQTT flooding scenarios; detection delay is decomposed into export, window-fill and inference terms (Section VII).

Three findings bound the claims. First, over ten seeds the hybrid has the lowest false-negative rate (FNR) on CIC-DDoS2019 at a 1% budget, and on CICIoT23 it matches the tree while the GRU alone misses 5.8% of attack windows. Second, the hybrid is not the most accurate model: an unconstrained tree exceeds it on CIC-DDoS2019, and trees and tree ensembles exceed it on CICIoT23, so its case rests on a short, readable rule for every decision and a measured footprint. Third, much of what these benchmarks appear to show does not survive the audits: Edge-IIoTset is saturated by duplicated records, none of the benchmarks, as used here, offers interleaved traffic in capture order, so the benefit of a window is that of pooling records of one class, and one CICIoT23 feature on which the tree relies is a probable artefact.

## II. Related Work

*Tightly coupled neuro-symbolic detectors.* Logic tensor networks embed first-order constraints in the training loss. Bizzarri *et al.* [13] add a satisfiability loss for intrusion detection in software-defined networks, and Onchis *et al.* [14], [15] show that logical regularisation reduces overfitting and data requirements; both lines of work report that cost grows with the number of axioms and classes. DeNSAINet [4] pairs a CNN with symbolic rules for DDoS detection in softwarised networks, with dataset-specific thresholds.

*Post-hoc and knowledge-driven explanation.* Almadhor *et al.* [16] extract symbolic rules from neural classifiers with SHAP and LIME, so the symbolic part operates after the decision. Kalutharage *et al.* [17] and Nerella *et al.* [18] combine autoencoders with attribution and expert knowledge graphs that are maintained by hand. Other designs add uncertainty quantification and open-set recognition [19], [20], federate a small CNN with a fuzzy-logic reasoner [21], shield a reinforcement-learning classifier with an SMT solver [22], stack graph networks, recurrent networks and federated learning [23], or learn graph representations without labels [24].

*Learning-based detection for OT and the edge.* Recurrent and convolutional detectors for industrial networks [7], [25], reinforcement-learning detectors [6], [26], [27] and lightweight CNNs for DDoS detection such as LUCID [28] address accuracy and, in part, inference cost. They are opaque, and few report latency, memory or energy on industrial hardware.

*Evaluation practice.* A separate literature documents how machine-learning results in security are distorted by data snooping and temporal bias [9]–[11], by labelling and design errors in the CIC benchmark family [29]–[32], by the base-rate fallacy [12], and by closed-world assumptions [33], and it formulates reporting guidance [34]. The methodology of Section V applies this guidance to windowed detectors on flow benchmarks.

*Relation to our earlier work.* Our survey [35] reviews AI techniques for the detection, mitigation and prevention of DDoS attacks in the industrial IoT. In [7] a lightweight deep model is pre-trained on CIC-DDoS2019 and adapted to CICIoT23 by sequential transfer learning, and in [6] a detector based on proximal policy optimisation is evaluated on the three benchmarks used here, with detection metrics and CPU inference latency. Both detectors are purely neural. The present paper asks what can be claimed for a detector whose every decision carries a readable rule, and it evaluates that detector under the stricter protocol of Section V.

*Positioning.* In Kautz's taxonomy [36] (see also [37]) the detector studied here is loosely coupled: separately trained neural and symbolic modules are reconciled at decision time, and the fusion is the weighted-sum rule of classifier combination [38]. The fusion rule is not new. The paper adds its selection under an explicit operational constraint, an audit of the rules that are actually learned, an evaluation that separates what the benchmarks support from what they do not, and a footprint measured on an industrial gateway. Where interpretability is required, a compact model that is read directly is preferable to a post-hoc explanation of an opaque one [39].

## III. System Model, Threat Model and Problem Statement

### A. Deployment

We adopt the Purdue reference architecture as used in NIST SP 800-82r3 [40]: basic control at Level 1, supervisory and site operations at Levels 2 and 3, and an industrial demilitarised zone (IDMZ, Level 3.5) between the plant and the enterprise network. Conduit traffic crosses the IDMZ aggregation switch, and the detector is attached there as a passive sensor on a mirror (SPAN) port or network tap (Fig. 1). It receives a copy of the traffic, reports alerts to a security operations centre, and has no write path to controllers. Two consequences follow. The sensor cannot delay or modify control traffic, so its latency requirement stems from the alerting path and not from PLC scan cycles. And it forms no part of a safety instrumented function: no claim under IEC 61508 or IEC 61511 is made, and an automated response driven by its alerts would need its own hazard assessment.

In terms of IEC 62443-3-3 [5] the sensor supports resource availability (FR 7): its alerts allow degraded-mode operation or upstream rate limiting to be activated before resource exhaustion propagates (SR 7.1, SR 7.2) and contribute to continuous monitoring (SR 6.2). In the NIST Cybersecurity Framework it implements the Detect function for the conduit. The task is binary, DoS/DDoS against benign traffic; other intrusion families are out of scope.

### B. Threat Model

The adversary is network-adjacent. It can inject traffic at the IDMZ switch, holds no PLC or broker credentials, and aims to exhaust resources rather than to manipulate process values. We consider six attack classes against the two protocols that most often cross the OT/IT boundary: Modbus/TCP read floods (function codes 3 and 4), Modbus/TCP write floods (function codes 16 and 23), TCP SYN floods on port 502, MQTT CONNECT floods, MQTT PUBLISH floods, and SlowITe [41], a low-rate attack on the MQTT keep-alive mechanism. The detector does not parse these protocols. It operates on flow records (packet-length and inter-arrival statistics, flag counts, packet and byte rates) of the kind exported by CICFlowMeter, and this representation is what links the public benchmarks of Section V to the testbed of Section VII. Whether statistics learned from benchmark floods carry over to Modbus and MQTT floods is examined in Section VII-D.

### C. Latency Quantities

Two quantities are kept apart throughout. The *per-window inference latency* $T_{\mathrm{inf}}$ is the time to scale one window, evaluate both branches and fuse their scores. We adopt a budget of $T_{\max} = 50$ ms for its 99th percentile, well below supervisory polling periods, which are of the order of a second and bound how quickly an alert can inform an operator. The *detection delay* $T_{\mathrm{det}}$ is the time from the first attack packet on the wire to the first alert. With flow-level sensing it decomposes as

$$T_{\mathrm{det}} = T_{\mathrm{exp}} + T_{\mathrm{fill}} + T_{\mathrm{inf}}, \tag{1}$$

where $T_{\mathrm{exp}}$ is the time until attack flows are exported, on connection teardown or on the timeouts of the exporter, and $T_{\mathrm{fill}}$ is the time until the scored window reflects the attack. At a flow rate $\lambda$ a window of $L$ records is renewed completely after $L/\lambda$, so $T_{\mathrm{fill}} \le L/\lambda$; a branch that reads only the newest record can respond to the first exported attack flow. Only $T_{\mathrm{inf}}$ is a property of the model, and it is never presented as the time to detect.

### D. Problem Statement

Let $\mathbf{z}_t \in \mathbb{R}^d$ be the scaled feature vector of the $t$-th flow record of a capture, $\mathbf{Z}_t = [\mathbf{z}_{t-L+1}, \ldots, \mathbf{z}_t]$ the window that ends at $t$, and $y_t \in \{0, 1\}$ the label of its last record. A detector is a score function $s : \mathbb{R}^{L\times d} \to [0, 1]$ with a threshold $\tau$; it raises an alert when $s(\mathbf{Z}_t) \ge \tau$. Writing TPR and FPR for its true- and false-positive rates, the design problem is

$$\max_{s\in\mathcal{S},\, \tau} \ \mathrm{TPR}(s, \tau) \quad \text{subject to} \quad \mathrm{FPR}(s, \tau) \le \beta, \tag{2}$$

over a model class $\mathcal{S}$ that meets three side conditions: the 99th percentile of $T_{\mathrm{inf}}$ on the target gateway is below $T_{\max}$, resident memory stays within the gateway's budget, and every alert comes with an explanation that an operator can read without access to the model. Problem (2) is the Neyman-Pearson formulation of classification [42]. It encodes the asymmetry of OT monitoring, in which a missed flood is costly while a sustained stream of false alerts exhausts the analyst, and unlike F1 its constraint does not depend on the class mix of the test set. Section IV specifies $\mathcal{S}$, and Section VI reports how far the constraint, imposed on validation data, holds on test data.

## IV. Hybrid Neural-Symbolic Detector

### A. Neural Branch

The neural branch is a two-layer GRU [43] with $h = 64$ hidden units and dropout 0.25, preceded by a layer normalisation of the input. For each decision the recurrence is run over the $L = 16$ vectors of $\mathbf{Z}_t$ from a zero initial state, and the final hidden state $\mathbf{h}_t$ is mapped to the score

$$p_{\mathrm{n}}(\mathbf{Z}_t) = \sigma\left(\mathbf{w}^\top \mathbf{h}_t + b\right), \qquad \sigma(u) = \frac{1}{1 + e^{-u}}. \tag{3}$$

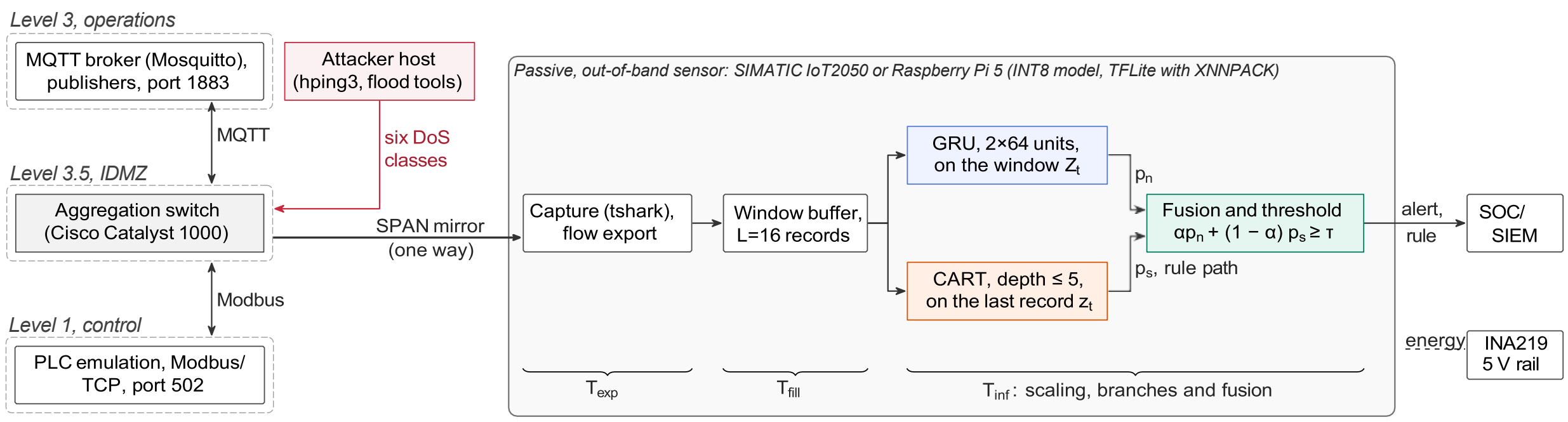

Fig. 1. Deployment, testbed and detection path. Left: the sensor is attached out of band to the mirror port of the IDMZ aggregation switch; the Modbus/TCP and MQTT endpoints of the testbed are software emulations. Right: processing on the gateway; an alert carries the root-to-leaf path of the tree, and braces mark the terms of the detection delay (1).

Restarting from zero for every window makes a decision depend on exactly $L$ records, which bounds both the memory of the detector and the influence of any single flow. A GRU is preferred to an LSTM for its smaller parameter count at equal width [25]. Training minimises the binary cross-entropy with positive-class weight $w_{\text{pos}} = N_{\text{neg}}/N_{\text{pos}}$ computed on the final training windows (2.33 on CIC-DDoS2019 and CICIoT23 after under-sampling, 4.78 on Edge-IIoTset). We use AdamW with a one-cycle schedule (peak learning rate $10^{-3}$, batch size 256) for at most 20 epochs, with early stopping on the validation loss; all hyperparameters are listed in the supplement.

## B. Symbolic Branch

The symbolic branch is a CART decision tree [44], grown with the Gini criterion and balanced class weights on the last record $\mathbf{z}_t$ of each training window, so that both branches predict for the same time index. Its score $p_s(\mathbf{z}_t)$ is the class-weighted attack fraction of the training samples in the leaf reached by $\mathbf{z}_t$. Depth is limited to five (two on Edge-IIoTset) with at least 100 samples per split and 40 per leaf (80 on Edge-IIoTset), so every decision is a conjunction of at most five threshold tests. Each threshold is mapped back to physical units by inverting the scaler of Section V-C. The tree is symbolic in this sense only: each of its paths is an auditable conjunction of predicates. It encodes no prior knowledge and performs no logical inference.

## C. Fusion and Operating Point

The hybrid score is the convex combination

$$p_h = \alpha\, p_n + (1-\alpha)\, p_s, \qquad \alpha \in [0,1], \tag{4}$$

and the class $\mathsf{S}$ of (2) is this one-parameter family. The loose coupling is a design choice. Each branch can be inspected, retrained or replaced without touching the other, and the rule set can be version-controlled independently of the network weights.

The pair $(\alpha, \tau)$ is selected on validation data by solving the empirical form of (2),

$$(\alpha_\beta, \tau_\beta) = \arg\max_{\alpha\in\mathsf{A},\,\tau} \mathrm{TPR}_v(\alpha,\tau) \quad \text{s.t.} \quad \mathrm{FPR}_v(\alpha,\tau) \le \beta, \tag{5}$$

where the subscript v denotes validation data, $\mathsf{A} = \{0, 0.025, \ldots, 1\}$ and $\beta \in \{0.1\%, 0.5\%, 1\%\}$. For each $\alpha$ the threshold is exact: with $n_0$ benign validation windows and $k = \lfloor \beta n_0 \rfloor$, $\tau$ is placed just above the $(k{+}1)$-th largest benign score, which is the smallest threshold whose validation FPR does not exceed $\beta$. Ties in TPR are broken in favour of the lower FPR. The GRU alone ($\alpha = 1$), the tree alone ($\alpha = 0$) and every baseline receive their operating points from the same rule, so all models are compared at points chosen in the same way. Each budget fixes a point on the ROC curve and thereby implies, for any assumed prevalence, a cost ratio between missed attacks and false alerts [45]; the budget itself is what an operator can state directly. Two properties of (5) matter for what follows.

*Remark 1 (nesting).* Because $0, 1 \in \mathsf{A}$, the validation TPR of the selected hybrid at budget $\beta$ is at least that of either branch at the same budget. On validation data the hybrid cannot lose to its branches. Nothing guarantees the same ordering on test data.

*Remark 2 (the budget in finite samples).* If the benign validation scores were $n_0$ independent draws from a continuous distribution, the population FPR of $\tau_\beta$ would be distributed as the $(k{+}1)$-th smallest of $n_0$ uniform variates, that is, as $\mathrm{Beta}(k{+}1,\ n_0{-}k)$, with mean $(k{+}1)/(n_0{+}1) \approx \beta$ and standard deviation close to $\beta(1-\beta)/n_0$. The budget is then met on average and exceeded with probability $\Pr\{\mathrm{Bin}(n_0, \beta) \le k\}$, about one half [46]. Overlapping windows make matters worse, because the effective $n_0$ is far below the nominal count (Section V-D), and so does any drift between validation and test traffic. A budget that holds with high probability requires the more conservative order statistic of [46]: $k$ is replaced by $k_\delta = \max\{j : \Pr\{\mathrm{Bin}(n_0,\beta) \le j\} \le \delta\}$, which bounds the probability of exceeding the budget by $\delta$. For $n_0 = 20{,}000$, $\beta = 1\%$ and $\delta = 5\%$, $k$ falls from 200 to 176 and the expected FPR from 1.005% to 0.885% (Fig. 2a). We keep the plain rule (5) and report the test FPR that is actually attained (Fig. 2b).

## D. Complexity and Footprint

With input dimension $d$ the neural branch has

$$P = 3h(d+h+2) + 3h(2h+2) + (h+1) + 2d \tag{6}$$

parameters, the last term being the normalisation layer: 46,815 for $d = 47$ (CIC-DDoS2019), 41,577 for $d = 20$ and 43,905 for $d = 32$, or 183 kB in 32-bit floating point and about a quarter of that with 8-bit weights. One decision costs $3Lh(d + 3h)$ multiply-accumulate operations, $0.73 \times 10^6$ for CIC-DDoS2019, plus at most five comparisons for the tree and two multiplications for (4). The tree and the fusion are therefore negligible in time and memory (4.3 kB serialised), and the hybrid inherits the cost of the GRU.

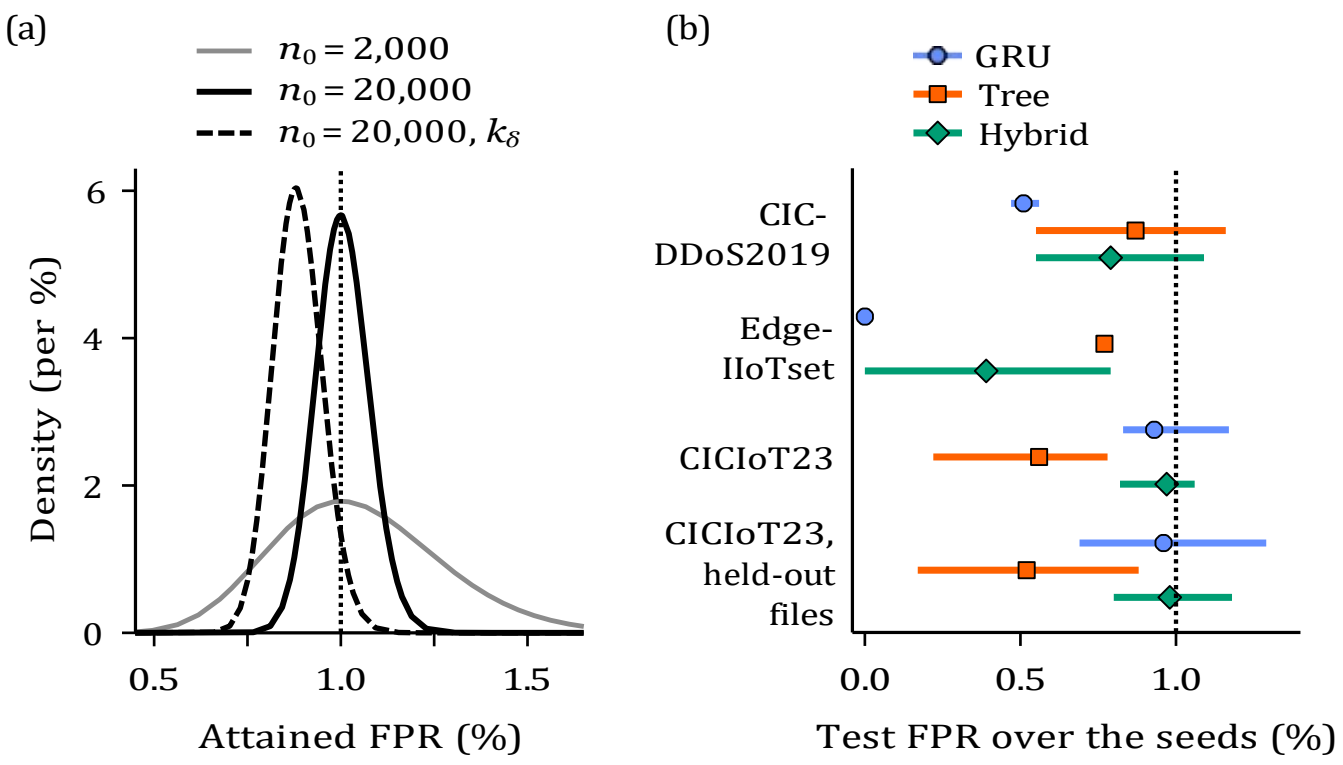

Fig. 2. The false-positive budget in finite samples. (a) Density of the population FPR attained by rule (5) at $\beta = 1\%$ with $n_0$ independent benign validation windows (Remark 2), and with the conservative $k_\delta$ for $\delta = 5\%$. (b) Test FPR over the ten seeds: mean (marker) and 2.5th to 97.5th percentile (bar); for the held-out files of CICIoT23, ten seeds times three folds. Dotted lines mark the budget. The day-held-out run of CIC-DDoS2019, in which the GRU is off the scale, is not shown.

## V. Evaluation Methodology

### A. Benchmarks and Task

Three public flow benchmarks are used (Table I). CIC-DDoS2019 [47] contains reflection and exploitation DDoS attacks captured on two days; we use the 17 per-attack files of a redistributed release, in which every attack class is a DoS variant. Edge-IIoTset [48] was recorded on a seven-layer IoT/IIoT testbed with Modbus/TCP and MQTT traffic and contains 14 attack classes; we use its selected-feature file, process it as one record sequence per traffic class, and keep the four flooding classes (TCP SYN, UDP, HTTP and ICMP) as the positive class. CICIoT23 [49] contains 33 attack classes in seven categories; we use 40 of its files, drawn at random with a fixed seed, and keep the DDoS and DoS categories. Records of the other attack families are removed from all partitions, because relabelling them as benign would introduce label noise. Attack windows are under-sampled to a 30% share in the training partition of CIC-DDoS2019 and CICIoT23 only. Validation and test partitions keep their native class mix, which is attack-dominated on two datasets, so accuracy is always read together with the Matthews correlation coefficient (MCC) and the per-class error rates.

### B. Benchmark Audits

Four properties of each benchmark are audited before any model is trained, because they determine what a windowed detector can be said to learn.

TABLE I
Benchmarks after the DoS restriction and results of the audits of Section V-B. Test windows refer to seed 42 (capture-held-out: fold 0).

| | CIC-DDoS2019 | Edge-IIoTset | CICIoT23 |
|---|---|---|---|
| Files used | 17 | 1 | 40 |
| Records | 431,371 | 2,219,201 | 10,974,530 |
| benign | 97,831 | 1,615,643 | 258,193 |
| DoS/DDoS (positive) | 333,540 | 337,977 | 9,890,490 |
| other families (removed) | 0 | 265,581 | 825,847 |
| Features, candidate / retained | 77 / 47 | 34 / 20 | 46 / 32 |
| Test windows, time-blocked | 64,170 | 292,886 | 469,029 |
| attack share | 33.6% | 17.3% | 97.4% |
| Test windows, capture-held-out | 306,051 | n/a | 541,439 |
| Record order of the data as used | label blocks* | per class† | random |
| Test records also in training | 2.0% | 99.8% | 1.4% |

*Contiguous label blocks in 16 files; the TFTP file is in random order.
†Processed as one sequence per traffic class; windows are single-class.

*Record order.* For each file we compare the fraction of adjacent records that share a label with its expectation under random order, $p^2 + (1 - p)^2$ for attack share $p$, and compute the mean lag-1 autocorrelation of the features. In 16 of the 17 CIC-DDoS2019 files the agreement is between 0.9989 and 1.0000, against 0.50 to 0.82 expected. Given the file sizes this corresponds to one or two label changes in 14 of these files and at most a handful in the other two: attack and benign records form contiguous blocks, with the benign block last (Fig. 3). The seventeenth file, TFTP, which holds 28% of all records, is in random order (0.695 against 0.695; autocorrelation 0.000). In every one of the 40 CICIoT23 files the agreement equals its random expectation to within 0.0003 and the autocorrelation is zero to within 0.002; the files also share one class mix (2.3% to 2.4% benign) and behave as random shards of a single pool. Windows are therefore label-homogeneous by construction on Edge-IIoTset and on 16 CIC-DDoS2019 files, and random samples of their file on CICIoT23 and on the TFTP file. As used here, none of the three benchmarks offers interleaved benign and attack traffic in capture order. What a window can contribute on these data is the pooling of $L$ records of one class, not the modelling of temporal dynamics, and the results below are read accordingly.

*Identifiers.* Identifier and shortcut-prone columns (flow identifiers, addresses, ports, timestamps, checksums, sequence and acknowledgement numbers, stream and transaction indices, payload text) are removed before feature selection, because in generated benchmarks they can encode the label [29], [32]. A depth-10 tree trained on the twelve numeric identifier columns of Edge-IIoTset alone reaches 99.90% F1, so these columns would have carried the label almost completely. After variance and correlation filtering (correlation above 0.98) on training records, 47, 20 and 32 features remain.

*Feature concentration.* Permutation importance, measured as the drop in test AUC, assigns the top feature 50.5%, 95.4% and 57.0% of the tree's importance mass on the three datasets and 21.2%, 49.1% and 69.0% of the GRU's. Removing the three features that rank highest for the GRU and refitting the tree changes its F1 from 99.39% to 99.12% on CIC-DDoS2019 and from 99.76% to 99.82% on CICIoT23, but from 80.9%

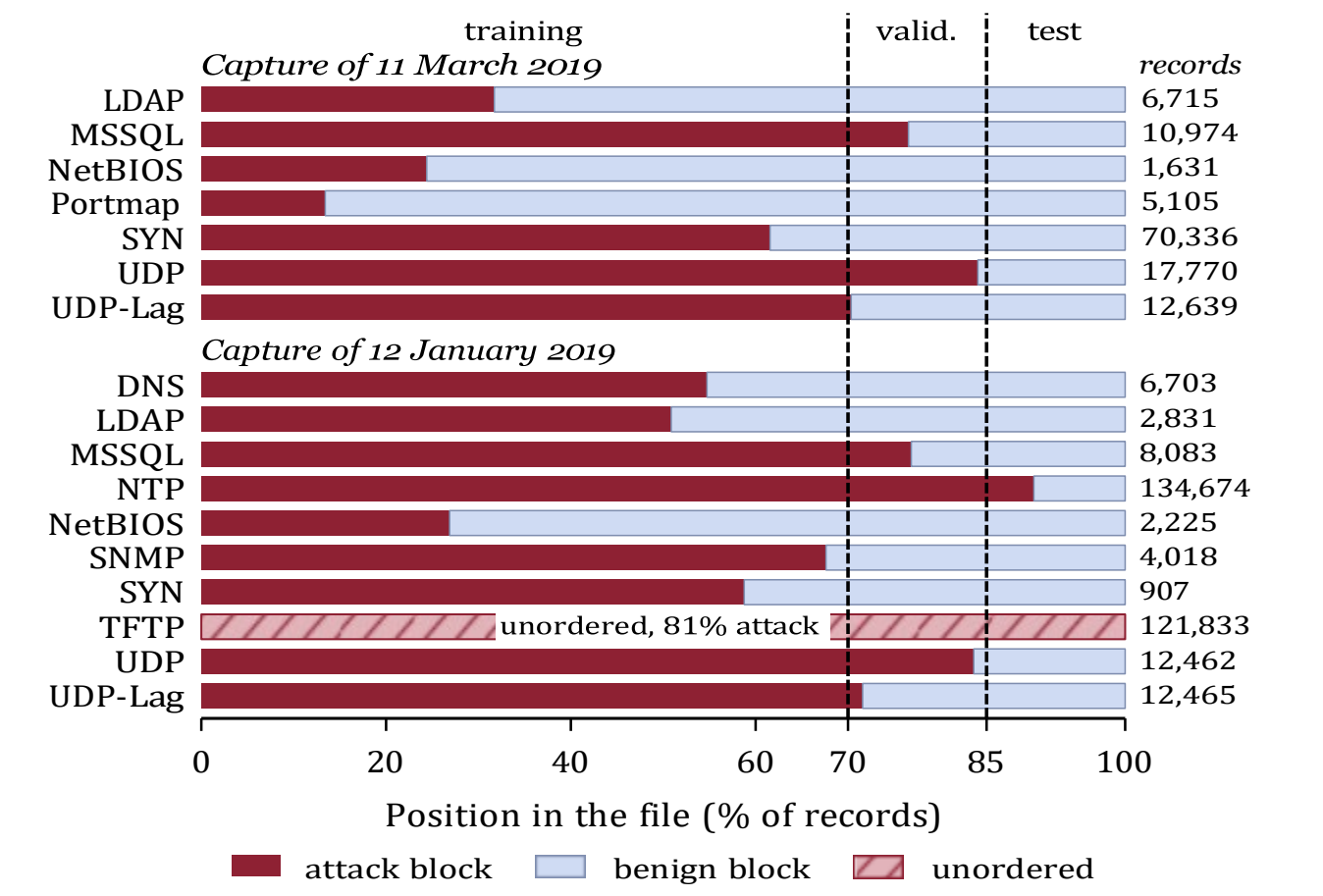

Fig. 3. Composition of the 17 CIC-DDoS2019 files in record order, as implied by the per-file counts and the order audit, with the cuts of the time-blocked scheme (dashed). Sixteen files consist of an attack block followed by a benign block; the TFTP file is unordered. Only the NTP file has attack records in its test block.

to 0% on Edge-IIoTset.

*Duplicates.* The share of test records whose retained feature vector also occurs among the training records is 2.0% on CIC-DDoS2019, 1.4% on CICIoT23 and 99.8% on Edge-IIoTset.

Taken together, Edge-IIoTset restricted to its flooding classes tests the recall of a few protocol fields seen in training more than detection, and it is used as a consistency check only. CICIoT23 measures what can be read from single records.

### C. Leakage-Controlled Protocol

The protocol follows from one requirement: a windowed detector must be tested on consecutive traffic that it has not seen in any form.

1) Source files are kept separate and their records in file order.
2) Each file is partitioned first. A window is then formed only from $L$ consecutive retained records that lie in one partition, so no window straddles a partition boundary or bridges a removed record. Windows are labelled by their last record and scored with stride one.
3) *Time-blocked scheme.* Each file is cut by record index into contiguous 70/15/15% training, validation and test blocks. The first $L$ records of each validation and test block are discarded as a purge gap, so that no validation or test window shares a record with a training window.
4) *Capture-held-out scheme.* Whole files are assigned to one partition, on CICIoT23 by grouped random assignment in three folds. On CIC-DDoS2019 the designation of the release is followed: the seven files of the capture day of 11 March 2019 (LDAP, MSSQL, NetBIOS, Portmap, SYN, UDP, UDP-Lag) supply training and validation data, and the ten files of 12 January 2019 the test data. This day-held-out split is also a test on unseen attack families, because 89% of the held-out attack records belong to NTP, TFTP, DNS and SNMP reflection attacks that do not occur on the training day. It is not a forward-in-time split. On Edge-IIoTset the scheme cannot be applied, because each sequence holds one class.
5) Under-sampling, the feature filters, the robust scaler $z = (x - \text{median})/\,\text{IQR}$ and the operating point are fitted on training or validation data only.

For reference we also run the record-level protocol that is common for these benchmarks, a stratified random 70/15/15 split of the records followed by window formation. After shuffling, a window holds unrelated records.

*Consequences for these benchmarks.* The audit of record order interacts with both schemes. On a file whose benign block comes last, the test block contains attack records only if less than 15% of the file is benign (Fig. 3). In CIC-DDoS2019 this holds for the NTP file alone, so the attack windows of the time-blocked test partition stem from the tail of the NTP attack block and from the unordered TFTP file (about 6,900 and 14,700 windows, respectively), and the test blocks of the other 15 files are benign. The class counts implied by this structure agree with those of the test partition to within 130 of 64,170 windows. The capture-held-out scheme is free of this effect, because every window of a held-out file is tested. On CICIoT23, whose files are random shards, both schemes amount to a random split of records.

### D. Dependence-Aware Uncertainty

Test windows are not independent: consecutive windows share $L - 1$ records, and windows of one file share its context. Binomial intervals on the nominal window count therefore understate the uncertainty and are not used. We report (i) a moving-block bootstrap [50] of the per-window outcomes in its circular form [51], in which a block of $b$ consecutive windows wraps around within its own run of consecutive windows, so that no block crosses a run and every window is equally likely to be drawn ($b \in \{16, 64, 256, 1024\}$, headline $b = 256$, $B = 2000$ resamples, percentile intervals); (ii) a cluster bootstrap over files [52], which adds between-capture variation; and (iii) the effective sample size

$$n_{\text{eff}} = n \left( \frac{w_{\text{W}}}{w_{\text{MBB}}} \right)^2 , \tag{7}$$

where $w_{\text{W}}$ and $w_{\text{MBB}}$ are the widths of the Wilson interval [53] and of the block-bootstrap interval for accuracy. Variability due to training is measured separately by repeating the complete pipeline (GRU, tree and operating-point selection) over seeds 42 to 51, crossed with the capture-held-out folds.

### E. Operating Points at Operational Prevalence

The test partitions are attack-dominated, whereas an OT conduit is almost always benign, and precision falls with prevalence [9], [12]. For an operating point with test rates TPR and FPR, the expected precision and the rate of false alerts at attack prevalence $\pi$ are

$$\text{Prec}(\pi) = \frac{\text{TPR}\,\pi}{\text{TPR}\,\pi + \text{FPR}\,(1 - \pi)}, \tag{8}$$

$$A(\pi) = \text{FPR}\,(1 - \pi)\, r_{\text{dec}}, \tag{9}$$

TABLE II
TEST PERFORMANCE FOR SEED 42 (TIME-BLOCKED, $\beta = 1\%$) WITH FALSE-POSITIVE (FP) AND FALSE-NEGATIVE (FN) WINDOW COUNTS.

| Dataset | Model | Acc | F1 | FPR | FNR | FP | FN |
|---|---|---|---|---|---|---|---|
| CIC-DDoS2019 | GRU | 99.56 | 99.35 | 0.56 | 0.19 | 240 | 42 |
| (42,623 benign, | Tree | 99.15 | 98.75 | 1.19 | 0.16 | 509 | 35 |
| 21,547 DoS) | Hybrid | 99.57 | 99.36 | 0.61 | 0.08 | 259 | 17 |
| Edge-IIoTset | GRU | 100.00 | 100.00 | 0.00 | 0.00 | 5 | 0 |
| (242,315 benign, | Tree | 94.26 | 80.92 | 0.77 | 29.53 | 1,873 | 14,932 |
| 50,571 DoS) | Hybrid | 100.00 | 100.00 | 0.00 | 0.00 | 5 | 0 |
| CICIoT23 | GRU | 93.99 | 96.82 | 0.97 | 6.14 | 117 | 28,067 |
| (12,019 benign, | Tree | 99.64 | 99.81 | 0.80 | 0.35 | 96 | 1,613 |
| 457,010 DoS) | Hybrid | 99.63 | 99.81 | 0.99 | 0.35 | 119 | 1,600 |

TABLE III
COMPLETE PIPELINE OVER TEN SEEDS (CICIOT23 CAPTURE-HELD-OUT: TEN SEEDS TIMES THREE FOLDS), $\beta = 1\%$; MEAN $\pm$ STANDARD DEVIATION IN %.

| Dataset, scheme | Model | F1 | FNR | FPR |
|---|---|---|---|---|
| CIC-DDoS2019, time-blocked | GRU | 99.37 ± 0.04 | 0.26 ± 0.04 | 0.51 ± 0.04 |
| | Tree | 99.06 ± 0.20 | 0.17 ± 0.01 | 0.87 ± 0.21 |
| | Hybrid | 99.19 ± 0.18 | 0.08 ± 0.01 | 0.79 ± 0.18 |
| CIC-DDoS2019, day-held-out | GRU | 95.62 ± 0.01 | 0.02 ± 0.02 | 45.29 ± 0.12 |
| | Tree | 98.37 ± 0.00 | 3.13 ± 0.00 | 0.40 ± 0.00 |
| | Hybrid | 98.37 ± 0.00 | 3.13 ± 0.00 | 0.40 ± 0.00 |
| Edge-IIoTset, time-blocked | GRU | 99.99 ± 0.00 | 0.00 ± 0.01 | 0.00 ± 0.00 |
| | Tree | 80.92 ± 0.00 | 29.53 ± 0.00 | 0.77 ± 0.00 |
| | Hybrid | 99.07 ± 0.97 | 0.01 ± 0.01 | 0.39 ± 0.41 |
| CICIoT23, time-blocked | GRU | 96.99 ± 0.44 | 5.82 ± 0.82 | 0.93 ± 0.12 |
| | Tree | 99.80 ± 0.04 | 0.38 ± 0.09 | 0.56 ± 0.18 |
| | Hybrid | 99.81 ± 0.04 | 0.36 ± 0.08 | 0.97 ± 0.09 |
| CICIoT23, capture-held-out | GRU | 97.06 ± 0.46 | 5.69 ± 0.86 | 0.96 ± 0.16 |
| | Tree | 99.79 ± 0.04 | 0.40 ± 0.08 | 0.52 ± 0.22 |
| | Hybrid | 99.80 ± 0.04 | 0.37 ± 0.08 | 0.98 ± 0.12 |

where $r_{\text{dec}}$ is the decision rate. We use $r_{\text{dec}} = 340$ flows/s, the sustained flow rate measured on the testbed (Section VII). The analytical values were checked against test sets resampled to the target prevalence, with attack windows drawn in contiguous blocks, and agree to within 0.1 percentage points.

## VI. RESULTS ON THE BENCHMARKS

### A. Setup

All benchmark experiments run in one version-pinned environment (Python 3.13.16, PyTorch 2.11.0, scikit-learn 1.6.1, XGBoost 3.4.1, one NVIDIA T4 GPU). Unless stated otherwise, values are test-set percentages under the time-blocked scheme at the operating point of (5) with $\beta = 1\%$, single-run analyses use seed 42, and Edge-IIoTset and CICIoT23 denote the DoS-restricted tasks. The baseline, depth-sweep and audit analyses refit the GRU and the tree; on CIC-DDoS2019 and CICIoT23 the refitted scores differ from the corresponding rows of Table II by amounts comparable to the seed-to-seed spread of Table III.

### B. Detection Performance

Table III reports the complete pipeline over ten seeds (percentile ranges are in the supplement) and Table II the seed-42 run in detail.

*CIC-DDoS2019.* The hybrid has the lowest FNR, 0.08 ± 0.01% against 0.26 ± 0.04% for the GRU and 0.17 ± 0.01% for the tree, and the percentile ranges over the seeds do not overlap. Its FPR (0.79 ± 0.18%) lies between the GRU's and the tree's and exceeds the 1% budget in some seeds (97.5th percentile 1.09%), as Remark 2 leads one to expect. Fig. 2b shows the same for every model: the upper percentiles of the tree and the hybrid on CIC-DDoS2019, and of the GRU and the hybrid on CICIoT23, lie above the budget, which is met on average and not in every run. Its mean F1 (99.19%) is below the GRU's (99.37%), because it spends more of the false-positive budget: the gain is a move along the trade-off as much as a better trade-off. In the seed-42 run the hybrid misses 17 of 21,547 attack windows where the GRU misses 42 and the tree 35, and raises 259 false alerts on 42,623 benign windows against 240 and 509. By Section V-C these attack windows are NTP and TFTP windows.

*CICIoT23.* The tree is already strong, the GRU misses 5.8% of attack windows at the same budget, and the hybrid, which selects $\alpha$ between 0.025 and 0.15, is indistinguishable from the tree in F1 and FNR. Fusion here protects against the weaker branch and adds nothing to the stronger one. It also shows a side effect of (5): the rule spends the whole budget whenever this buys any validation recall, however little, and the hybrid's FPR is 0.97% where the tree, whose scores are discrete, stays at 0.56%. The weakness of the GRU is consistent with the audit: on unordered files the 15 earlier records of a window carry no information about the label of the last one.

*Edge-IIoTset.* The GRU and the hybrid make five errors in 292,886 windows in the seed-42 run, whereas the depth-2 tree misses 29.5% of attacks. Given the audits we draw no conclusion from this. Over ten seeds the hybrid (F1 99.07 ± 0.97%) is in fact worse than the GRU (99.99%), because in four seeds the validation-selected weight moved towards the weak tree, a consequence of selecting on near-perfect validation scores.

The selected weight shows which branch carries the decision. Over the seeds it ranges from 0.05 to 0.975 on CIC-DDoS2019, where both branches reach high recall and the validation objective is nearly flat in $\alpha$, from 0.725 to 1.0 on Edge-IIoTset, and from 0.025 to 0.15 on CICIoT23 (operating points and validation surfaces are in the supplement). Fig. 4 shows the test ROC curves on a logarithmic FPR axis, the region in which an operator works. On CIC-DDoS2019 both branches lose recall for FPR under $10^{-4}$, the GRU more than the tree, whereas the fused score does not; this region corresponds to fewer than five of the 42,623 benign test windows, so the difference is suggestive only. On CICIoT23 the fused curve stays close to the tree's, and on Edge-IIoTset it coincides with the GRU's. The hybrid's AUC exceeds 99.99% on CIC-DDoS2019 and Edge-IIoTset and is 99.92% on CICIoT23.

### C. Effect of the Protocol and Held-Out Captures

Table IV compares protocols for the hybrid. With the same positive class, the choice between a record-level and a time-blocked split moves accuracy by less than 0.1 points on CIC-DDoS2019 and CICIoT23. Their classes are separable from single records, so the record-level protocol does not inflate these headline numbers by much. It matters for two other reasons. First, it changes what is measured: after shuffling, a window holds unrelated records, and on Edge-IIoTset the hybrid loses 12.7 points of recall under the record-level protocol because the GRU's windows lose their structure. Second, only held-out captures expose transfer effects.

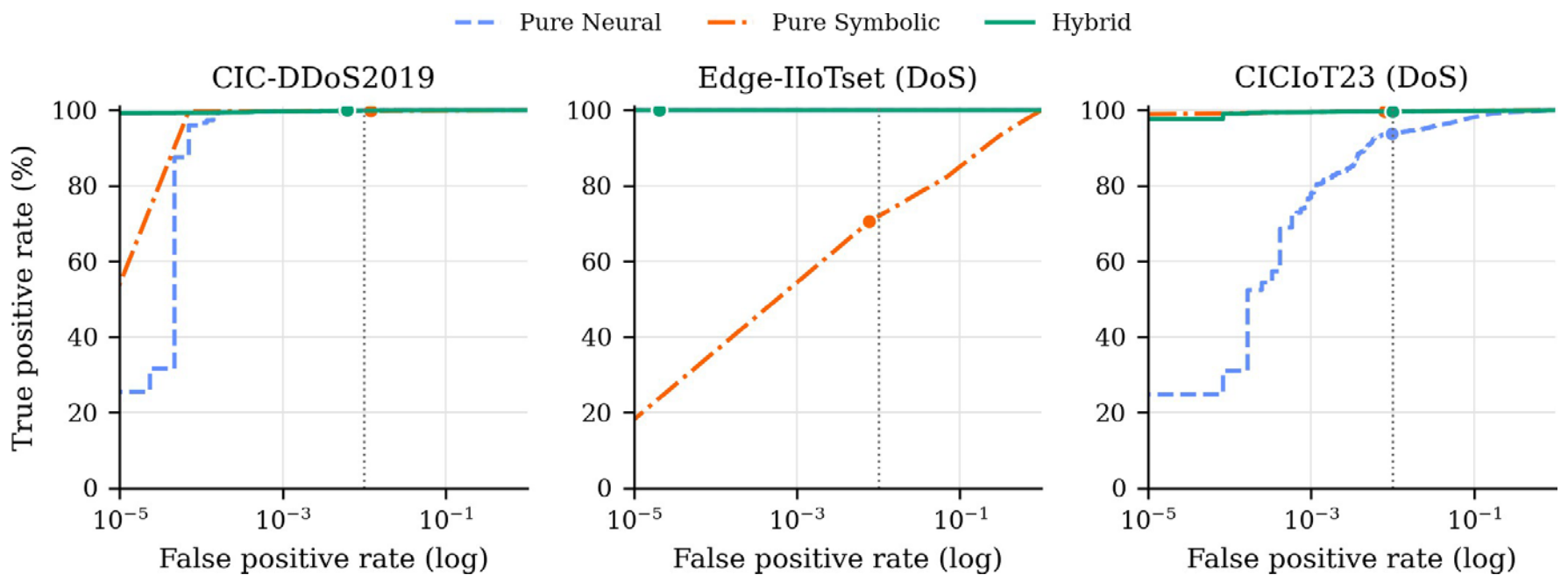

Fig. 4. Test ROC curves on a logarithmic false-positive axis (time-blocked scheme, seed 42). Dots mark the operating points of (5) and the dotted line the 1% budget.

TABLE IV
EFFECT OF THE EVALUATION PROTOCOL ON THE HYBRID (SEED 42, $\beta = 1\%$). F1 IS NOT COMPARABLE ACROSS ROWS, BECAUSE THE CLASS MIX OF THE TEST PARTITION DIFFERS.

| Dataset | Protocol | Acc | F1 | FPR | FNR | MCC |
|---|---|---|---|---|---|---|
| CIC-DDoS2019 | Record-level | 99.54 | 99.70 | 1.04 | 0.29 | 0.99 |
| | Time-blocked | 99.57 | 99.36 | 0.61 | 0.08 | 0.99 |
| | Day-held-out | 97.33 | 98.37 | 0.39 | 3.13 | 0.91 |
| Edge-IIoTset | Record-level, all attacks | 89.80 | 77.67 | 1.00 | 34.80 | 0.74 |
| | Record-level | 96.98 | 90.92 | 0.99 | 12.70 | 0.89 |
| | Time-blocked | 100.00 | 100.00 | 0.00 | 0.00 | 1.00 |
| CICIoT23 | Record-level, all attacks | 98.72 | 99.34 | 1.13 | 1.28 | 0.80 |
| | Record-level | 99.62 | 99.81 | 1.03 | 0.36 | 0.93 |
| | Time-blocked | 99.63 | 99.81 | 0.99 | 0.35 | 0.93 |
| | Capture-held-out | 99.61 | 99.80 | 0.93 | 0.37 | 0.93 |

TABLE V
DEPENDENCE-AWARE 95% INTERVALS FOR THE HYBRID (TIME-BLOCKED, SEED 42, $\beta = 1\%$): MOVING-BLOCK BOOTSTRAP (MBB, $b = 256$) AND CLUSTER BOOTSTRAP OVER FILES (CB).

| Dataset ($n$ / $n_{\text{eff}}$) | CI | Acc (%) | FPR (%) | FNR (%) |
|---|---|---|---|---|
| CIC-DDoS2019 | MBB | [99.48, 99.65] | [0.48, 0.77] | [0.04, 0.12] |
| (64,170 / 22,935) | CB | [99.18, 99.86] | [0.16, 1.78] | [0.00, 0.12] |
| CICIoT23 | MBB | [99.55, 99.70] | [0.42, 1.74] | [0.28, 0.43] |
| (469,029 / 25,437) | CB | [99.61, 99.65] | [0.81, 1.18] | [0.33, 0.37] |

On CICIoT23 the capture-held-out results cannot be told apart from the time-blocked ones (Table III), as expected for random shards. On CIC-DDoS2019 the day-held-out split changes the picture. The tree transfers: its FPR is 0.40% and its FNR rises from 0.17% to 3.13%, although 89% of the held-out attack records come from families that are absent from the training day, which suggests that packet-length and timing statistics of the kind tested by the rules of Section VI-F are shared by the unseen reflection attacks. The GRU keeps 99.98% recall, but its test FPR at the 1% validation budget is $45.3 \pm 0.1\%$. Rule (5), applied to validation data of the training day, selects $a = 0$ in all ten seeds, and the hybrid coincides with the tree (accuracy 97.33%, MCC 0.91). This fall-back is a property of the validation data. It is not a response to the shift, which (5) cannot see, and we do not claim that fusion protects against distribution shift.

The order audit suggests a specific reading of the GRU's failure. The unordered TFTP file belongs to the held-out day and contains 22,916 of its 51,404 benign records (44.6%). In that file a window labelled benign by its last record consists, on average, of 81% attack records in its other positions, whereas on the training day nearly every window is label-homogeneous. A recurrent model trained on such windows has no reason to disregard the attack content of a mixed window, and an FPR of 45.3% is what results if the GRU flags every benign-labelled TFTP window and about 1.3% of the remaining benign windows. We therefore read the 45% figure as a disagreement between window content and last-record labels on a shuffled file more than as drift between the days. The tree reads the last record only and is immune to the effect by construction. Selection is fragile on this split for a second reason: with benign blocks last, the validation blocks of the training day contain attack windows from one file only (about 270 windows of the UDP file), so $a$ is chosen on a narrow sample. In either reading, a detector of this kind should be validated, and its operating point chosen, on held-out captures that resemble the deployment traffic.

## D. Uncertainty

The dependence between windows is substantial (Table V). On CICIoT23 the block-bootstrap interval for accuracy is 4.3 times as wide as the Wilson interval, so by (7) the 469,029 test windows carry the information of about 25,000 independent ones, and the FPR interval spans 0.42% to 1.74% where the Wilson interval gives 0.83% to 1.18%. On CIC-DDoS2019 the inflation is smaller ($n_{\text{eff}} = 22{,}935$ of 64,170). The cluster interval for its FPR is wider still (0.16% to 1.78%), because the false alerts concentrate in a few of the 17 files. Under the day-held-out scheme the dependence is extreme: the block-bootstrap interval for the accuracy of the hybrid is [96.58, 98.04]%, which corresponds to $n_{\text{eff}} \approx 1{,}900$ for 306,051 windows. On Edge-IIoTset the intervals are degenerate at the ceiling ($n_{\text{eff}} = 119{,}405$ of 292,886).

TABLE VI
EXPECTED PRECISION (%) AND FALSE ALERTS PER HOUR FROM (8) AND (9) FOR THE HYBRID, WITH $r_{dec}$ = 340 FLOWS/S, AND THE TEST FNR OF EACH POINT.

| | | Prec($\pi$) at $\pi$ = | | | $A$(1%) | FNR |
|---|---|---|---|---|---|---|
| **Dataset** | $\beta$ | 0.1% | 1% | 5% | (h$^{-1}$) | (%) |
| CIC-DDoS2019 | 1% | 14.13 | 62.42 | 89.64 | 7,363 | 0.08 |
| | 0.1% | 79.49 | 97.51 | 99.51 | 313 | 0.11 |
| Edge-IIoTset | 1% | 97.98 | 99.80 | 99.96 | 25 | 0.00 |
| | 0.1% | 76.62 | 97.07 | 99.42 | 370 | 0.00 |
| CICIoT23 | 1% | 9.15 | 50.41 | 84.12 | 11,998 | 0.35 |
| | 0.1% | 44.38 | 88.95 | 97.67 | 1,512 | 0.50 |

The conclusions of Section VI-B survive these intervals on CIC-DDoS2019, where the hybrid's FNR interval lies below the point estimates of the GRU and the tree. They do not support a separation of the hybrid from the tree on CICIoT23, whose intervals overlap almost completely. The block-length sweep changes the CIC-DDoS2019 FPR interval only from [0.51, 0.72] at $b$ = 16 to [0.40, 0.85] at $b$ = 1024, and strides of 4 and 16 give intervals consistent with those of stride one.

### E. Precision and Alert Load at Operational Prevalence

Table VI applies (8) and (9) to the selected operating points. The 1% budget is too loose for unattended use: on CIC-DDoS2019 it yields a precision of 14% at a prevalence of 0.1% and about 7,400 false alerts per hour at 340 decisions per second. The stricter budget $\beta$ = 0.1% reduces the load to 313 alerts per hour and raises the precision at 1% prevalence from 62% to 97.5%, for an FNR of 0.11%. It is the operating point we recommend on these data. On CICIoT23 the same change gives 1,512 alerts per hour at an FNR of 0.50%. On the saturated Edge-IIoTset the test FPRs lie far below both budgets and even change order (25 and 370 false alerts per hour), one more reason not to read that dataset quantitatively. All rates assume one decision per flow; aggregation per source or time slot, as an operator would apply, reduces them further. Fig. 5 extends the table to the whole range of prevalence. At the 1% budget the precision of every model on CIC-DDoS2019 and CICIoT23 is below 20% once fewer than one flow in a thousand belongs to an attack, and the choice of budget matters far more than the choice of model.

### F. Learned Rules, Stability and Complementarity

Every root-to-leaf path of the trees is exported with its thresholds in physical units. Table VII lists the rules with the largest support.

*CIC-DDoS2019.* The root split is the minimum forward packet length at 79 B in all ten seeds. The main attack rule (minimum forward packet length above 79 B, mean packet size above 343 B, minimum inter-arrival time of at most 109 $\mu$s, bounded length variation) covers 21,444 of the 21,547 attack test windows at a purity of 99.99%. It describes large, uniform datagrams at sub-millisecond spacing, the signature of UDP reflection and amplification, whereas benign flows in these captures are dominated by short TCP segments.

*Edge-IIoTset.* The depth-2 tree splits on the TCP flag byte. Its attack leaf requires tcp.flags $\leq$ 3 and no connection-request marker, a condition met by records that carry no TCP header. The tree detects 35,639 attack windows; by count, the test block holds 35,638 UDP- and ICMP-flood windows and 14,933 TCP SYN- and HTTP-flood windows, and the tree misses 14,932. The rule thus isolates the two non-TCP floods and fails on the two TCP-based ones, whose flag patterns it cannot separate from benign TCP traffic at this depth, which accounts for its FNR of 29.5%.

*CICIoT23.* The root split on the RST count is plausible, but the attack rules then require the `IAT` feature to lie between $2.25 \times 10^7$ and $1.33 \times 10^8$, and they have a purity of 100% on 451,226 test windows. Such values are far too large to be inter-arrival times of flood packets. The feature behaves like an index of capture time, and we flag it as a probable artefact of how the release was generated, of the kind documented for this benchmark family [32]. The CICIoT23 results should be read with this caveat.

*Stability.* Over the ten seeds the mean Jaccard similarity of the five most important features of the trees is 0.72 on CIC-DDoS2019 (minimum 0.43) and 0.72 on Edge-IIoTset, with the same root feature and threshold in every seed. On CICIoT23 it is 0.46: the root is the RST count in eight seeds (median threshold 1.20) and the URG count in two, so the lower levels of that tree should not be read as stable explanations.

*Complementarity.* From the cross-classification of the windows that the GRU and the hybrid decide correctly, fusion corrects 31 of the GRU's 282 errors on CIC-DDoS2019 and turns 25 of its 63,888 correct decisions into errors. The block-bootstrap intervals of the two rates, [0.033, 0.068]% and [0.023, 0.056]% of all windows, overlap. The net effect on the error count is therefore not distinguishable from zero, and the FNR gain of Table II is bought with false alerts. On CICIoT23 fusion corrects 26,523 of the GRU's 28,184 errors (94.1%) and breaks 58 correct decisions, a net gain of 5.6% of all windows. McNemar-type tests are not used, because they assume independent trials.

*Worked example.* A CIC-DDoS2019 attack window receives $p_n$ = 0.006, so the GRU alone misses it. The tree follows the path minimum forward packet length $\leq$ 79 B, mean flow inter-arrival time $\leq$ 2.45 s (here 20 $\mu$s), ACK flag set, initial forward window $\leq$ 7016 B (here 5840 B) to a pure attack leaf, and the fused score $0.975 \times 0.006 + 0.025 \times 1 = 0.031$ exceeds $\tau$ = 0.019. The alert is raised and the path is its explanation. Two further traced windows, one of them a false alert caused by the tree, are given in the supplement.

### G. Baselines, Ablations and the Price of Interpretability

Table VIII compares the hybrid and the GRU with classical and lightweight neural models. Every model sees the same length-16 windows, the same split and the same rule (5). Logistic regression, a decision tree, a random forest [54], XGBoost [55] and a multilayer perceptron (MLP) receive six summaries of each feature over the window (mean, standard deviation, minimum, maximum, last value and linear slope).

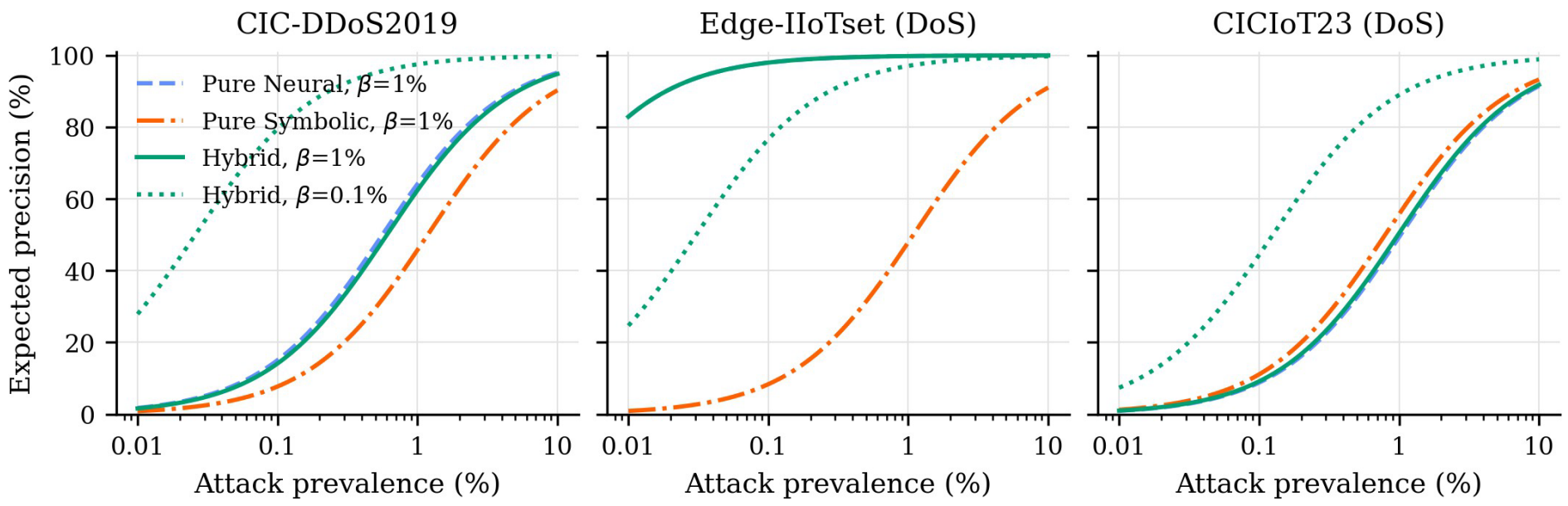

Fig. 5. Expected precision against attack prevalence from (8) for the three models at $\beta = 1\%$ and for the hybrid at $\beta = 0.1\%$, computed from the test TPR and FPR of the seed-42 run.

TABLE VII
HIGHEST-SUPPORT RULES OF THE TREES OF THE SEED-42 RUN, WITH THRESHOLDS IN PHYSICAL UNITS, AND THEIR SUPPORT AND PURITY ON TRAINING AND TEST WINDOWS. FEATURE NAMES ARE THOSE OF THE RELEASES; THE UNITS OF THE CICIOT23 FEATURES ARE AS DOCUMENTED FOR THAT RELEASE.

| **Dataset** | **Class** | **Rule** | **Train: support / purity** | **Test: support / purity** |
|---|---|---|---|---|
| CIC-DDoS2019 | DoS | Fwd Pkt Len Min > 79 B ∧ Avg Pkt Size > 343 B ∧ Flow IAT Min ≤ 109 µs ∧ Pkt Len Std ≤ 186 B ∧ Pkt Len Max > 300 B | 10,224 / 1.000 | 21,444 / 1.000 |
| | benign | Fwd Pkt Len Min ≤ 79 B ∧ Flow IAT Mean ≤ 2.45 s ∧ ACK = 0 ∧ Flow IAT Min ≤ 171 ms | 23,264 / 0.999 | 32,872 / 0.999 |
| | benign | Fwd Pkt Len Min ≤ 79 B ∧ Flow IAT Mean ≤ 2.45 s ∧ ACK = 1 ∧ Init Fwd Win > 7016 B ∧ Flow Pkts/s ≤ 876 | 6,046 / 0.999 | 7,939 / 0.999 |
| Edge-IIoTset | DoS | tcp.flags ≤ 3 ∧ tcp.connection.syn = 0 | 173,783 / 0.959 | 37,512 / 0.950 |
| | benign | 3 < tcp.flags ≤ 19 | 782,343 / 0.984 | 168,559 / 0.981 |
| CICIoT23 | DoS | rst_count ≤ 1.20 ∧ flow_duration ≤ 14.7 ∧ $2.25 \times 10^7$ < IAT ≤ $1.33 \times 10^8$ ∧ syn_count ≤ 0.005 | 16,722 / 1.000 | 320,466 / 1.000 |
| | DoS | as above with syn_count > 0.005 | 6,899 / 1.000 | 130,760 / 1.000 |
| | benign | rst_count > 1.20 ∧ Magnitude > 10.53 ∧ urg_count > 1.80 ∧ fin_count ≤ 0.005 ∧ Protocol Type > 3.65 | 52,922 / 0.999 | 12,369 / 0.922 |

A one-dimensional CNN in the style of LUCID [28] reads the raw window; it has two convolutional layers of 32 filters with global max pooling and 7,681 parameters on CIC-DDoS2019, one sixth of the GRU. Two ablations bound the contribution of the window: the GRU evaluated on windows whose records are shuffled, and a GRU trained with $L = 1$.

Three findings stand out. First, the hybrid is not the most accurate model. On CIC-DDoS2019 an unconstrained CART on the last record reaches 99.77% F1 against 99.36%, on CICIoT23 the random forest and XGBoost reach 99.99% against 99.81%, and on Edge-IIoTset the hybrid ties with the GRU and the CNN. Second, the window contributes little beyond pooling. With $L = 1$ the F1 of the GRU falls from 100% to 92.05% on Edge-IIoTset, where windows are single-class, from 99.29% to 98.89% on CIC-DDoS2019, and not at all on CICIoT23 (97.33% against 97.13%). Shuffling the window collapses the GRU on CIC-DDoS2019 (58.44%) and CICIoT23 (2.32%); because the label of a window is that of its last record, this shows that the GRU reads the most recent record, not that it models order. The CNN, whose pooling discards position, cannot single out that record, which is consistent with its FNR of 22% on the two datasets that contain unordered files. Third, at the common operating point several baselines exceed the 1% budget on test data (logistic regression at 1.33% and XGBoost at 1.02% on CICIoT23, the MLP at 1.02% on CIC-DDoS2019), while the hybrid stays within it in this run. Near-ceiling scores on this benchmark family separate methods only weakly [32].

Table IX prices interpretability on CIC-DDoS2019. An unconstrained tree gains 0.38 F1 points over the depth-5 configuration at four times the leaves (85 against 20) and a mean path of 8.1 tests. The ensembles reach similar accuracy at 0.37 MB (XGBoost) and 1.9 MB (random forest), and an explanation would require reading hundreds of trees. All of these models fit the memory of the gateway, so the choice among them is decided by explanation length and error profile, not by size. We did not measure the ensembles on the gateway.

### H. Sensitivity

*Window length.* $L \in \{1, 4, 8, 16, 32, 64\}$ is swept on a common set of test windows, those with at least 63 records of history (supplement, Table S9). On CIC-DDoS2019 the hybrid is insensitive to $L$ (F1 from 99.08% to 99.33%). On Edge-IIoTset F1 rises from 92.5% at $L = 1$ to 100% at $L = 16$ and 32 and falls at $L = 64$. On CICIoT23 F1 is flat, as expected for unordered files; the common test subset there holds only about 300 benign windows, so its FPR is coarse (1.6% to 3.2%) and is not comparable with Table II. $L = 16$ is within 0.1 F1 points of the best value on every dataset and fills in 47 ms at the testbed flow rate.

*Capacity and stride.* Hidden sizes of 32, 64 and 128 and depths of one and two layers change the hybrid's F1 by at most 0.07 points on CIC-DDoS2019 and 0.01 points on CICIoT23. On Edge-IIoTset four of the six configurations stay at or above 99.98% and two drop to 98.2% and 98.9%, because the selected weight moved towards the tree. Strides of 1, 4

TABLE VIII
GRU, HYBRID AND BASELINES ON IDENTICAL LENGTH-16 WINDOWS, THE SAME TIME-BLOCKED SPLIT AND THE SAME OPERATING-POINT RULE ($\beta$ = 1%, SEED 42; VALUES IN %). ROWS OTHER THAN THE HYBRID ARE FITTED IN THE BASELINE ANALYSIS (SECTION VI-A).

| | CIC-DDoS2019 | | | | Edge-IIoTset | | | | CICIoT23 | | | |
|---|---|---|---|---|---|---|---|---|---|---|---|---|
| **Model (input)** | F1 | AUC | FNR | FPR | F1 | AUC | FNR | FPR | F1 | AUC | FNR | FPR |
| Hybrid (window and last record) | 99.36 | 100.00 | 0.08 | 0.61 | 100.00 | 100.00 | 0.00 | 0.00 | 99.81 | 99.92 | 0.35 | 0.99 |
| GRU (window) | 99.29 | 99.99 | 0.20 | 0.62 | 100.00 | 100.00 | 0.00 | 0.00 | 97.13 | 99.33 | 5.55 | 0.90 |
| GRU (shuffled window) | 58.44 | 97.41 | 58.34 | 0.46 | 99.39 | 100.00 | 0.07 | 0.24 | 2.32 | 55.02 | 98.83 | 0.91 |
| GRU ($L$ = 1) | 98.89 | 99.94 | 0.32 | 0.97 | 92.05 | 96.82 | 12.83 | 0.46 | 97.33 | 99.53 | 5.17 | 0.97 |
| 1D-CNN (window) | 87.56 | 97.84 | 22.01 | 0.07 | 100.00 | 100.00 | 0.00 | 0.00 | 87.42 | 99.12 | 22.33 | 0.84 |
| Logistic regression (window summaries) | 99.25 | 99.98 | 0.08 | 0.73 | 98.42 | 100.00 | 3.10 | 0.00 | 99.94 | 99.98 | 0.08 | 1.33 |
| Decision tree (window summaries) | 99.53 | 99.71 | 0.23 | 0.35 | 99.40 | 99.41 | 1.18 | 0.00 | 99.92 | 99.90 | 0.15 | 0.05 |
| Random forest (window summaries) | 99.04 | 99.97 | 0.07 | 0.95 | 99.86 | 100.00 | 0.00 | 0.06 | 99.99 | 100.00 | 0.00 | 0.91 |
| XGBoost (window summaries) | 99.20 | 100.00 | 0.06 | 0.79 | 99.94 | 100.00 | 0.05 | 0.01 | 99.99 | 100.00 | 0.00 | 1.02 |
| MLP (window summaries) | 98.94 | 99.94 | 0.11 | 1.02 | 99.66 | 100.00 | 0.66 | 0.00 | 99.97 | 99.99 | 0.03 | 0.91 |
| CART (last record), depth 5 | 99.39 | 99.90 | 0.18 | 0.53 | 87.60 | 96.95 | 20.36 | 0.46 | 99.76 | 99.87 | 0.47 | 0.50 |
| CART (last record), unconstrained | 99.77 | 99.87 | 0.07 | 0.20 | 93.53 | 96.83 | 10.22 | 0.46 | 99.93 | 99.91 | 0.14 | 0.04 |

TABLE IX
DETECTION, SERIALISED SIZE AND EXPLANATION LENGTH ON CIC-DDOS2019 (TIME-BLOCKED, $\beta$ = 1%, SEED 42). THE TREE ROWS ARE FITTED IN THE DEPTH SWEEP.

| Model | F1 (%) | FPR (%) | Size (kB) | Leaves / mean path |
|---|---|---|---|---|
| CART, depth 2 | 99.37 | 0.46 | 1.7 | 4 / 2.0 |
| CART, depth 3 | 99.07 | 0.77 | 2.4 | 8 / 3.0 |
| CART, depth 5 (as symbolic branch) | 99.39 | 0.53 | 4.3 | 20 / 5.0 |
| CART, depth 8 | 99.58 | 0.30 | 6.1 | 31 / 6.8 |
| CART, unconstrained | 99.77 | 0.20 | 14.7 | 85 / 8.1 |
| Logistic regression | 99.25 | 0.73 | 10.2 | n/a |
| MLP | 98.94 | 1.02 | 550 | n/a |
| XGBoost | 99.20 | 0.79 | 367 | 300 trees |
| Random forest | 99.04 | 0.95 | 1,895 | 100 trees |
| GRU (FP32) | 99.29 | 0.62 | 183 | n/a |
| Hybrid (GRU and depth-5 CART) | 99.36 | 0.61 | 187 | 20 / 5.0 |

TABLE X
PER-WINDOW INFERENCE PROFILE OF THE DEPLOYED INT8 MODEL OVER 1,000 BATCH-ONE INFERENCES. THE WORKSTATION COLUMN IS THE FP32 REFERENCE; THE LAST ROW RELATES THE 99TH PERCENTILE TO THE 50 MS BUDGET.

| Metric | IoT2050 | Pi 5 | i7-11800H |
|---|---|---|---|
| Mean latency (ms) | 2.9 | 0.6 | 0.62 |
| Median latency (ms) | 2.7 | 0.55 | 0.60 |
| 95th percentile (ms) | 5.1 | 1.1 | 0.74 |
| 99th percentile (ms) | 6.8 | 1.6 | 0.95 |
| Maximum (ms) | 9.7 | 2.4 | 1.6 |
| Throughput, one thread (inferences/s) | 345 | 1,670 | 1,610 |
| Throughput, four threads (inferences/s) | 1,100 | 5,200 | n/a |
| Peak resident memory (MB) | 31 | 28 | 340 |
| Idle power (W) | 2.05 | 2.74 | 11.5 |
| Active power (W) | 3.25 | 4.90 | 27.0 |
| Net energy per inference (mJ) | 3.5 | 1.3 | n/a |
| Headroom at the 99th percentile | 7.4× | 31× | 53× |

and 16 change F1 by at most 0.12 points on CIC-DDoS2019 and CICIoT23.

## VII. Validation on an Industrial Edge Gateway

### A. Platforms and Deployed Artefact

The detector is deployed on a Siemens SIMATIC IoT2050 industrial gateway (TI AM6548, four Arm Cortex-A53 cores, 2 GB of memory) and, for reference, on a Raspberry Pi 5 (four Cortex-A76 cores, 8 GB). Both are powered from a regulated 5 V bench supply. The model runs as an INT8 TensorFlow Lite graph with the XNNPACK delegate [56] at batch size one, after a warm-up of at least 100 inferences. A workstation with an Intel Core i7-11800H provides an FP32 reference.

*Provenance.* The artefact on the gateway is the INT8 export of the CIC-DDoS2019 detector that was built for the first deployment of this system; its feature schema is the one produced by the flow extractor on the gateway. Its architecture is that of Section IV (Fig. 1): a two-layer GRU with 64 hidden units and $d$ = 47 inputs, 46,815 parameters by (6), fused with a CART tree of depth at most five, with no further components. It was trained with a record-level split, before the protocol of Section V-C was adopted, and the models of Section VI have so far been quantised and evaluated off the device only. This matters differently for the two kinds of measurement below. Latency, memory and energy are determined by the architecture, the arithmetic and the runtime and not by the weights, so Table X characterises the deployed network on this hardware whatever its training. The detection rates of Section VII-D do depend on the weights. They are reported as a functional check of the complete on-gateway path under OT-protocol floods and not as an evaluation of the models of Section VI.

### B. Effect of Quantisation

Before timing is discussed we check that 8-bit arithmetic does not change the decisions of the models of Section VI. Post-training dynamic quantisation of the GRU and of the output layer to 8-bit weights in PyTorch, with the FP32 operating point left unchanged, gives a per-window agreement with the FP32 hybrid of 99.97%, 100.00% and 99.99% on the three test partitions. Accuracy and F1 change by less than 0.01 points. The test FPR moves from 0.608% to 0.612% on CIC-DDoS2019 and from 0.965% to 1.082% on CICIoT23 (values recomputed on a CPU), so on the latter the operating point should be selected again after quantisation. The TensorFlow Lite conversion used on the gateway is a different toolchain, and the same comparison remains to be repeated for it.

### C. Latency, Memory, Energy and Capacity

Table X reports the profile. The IoT2050 needs 2.9 ms per window on average and 6.8 ms at the 99th percentile, the

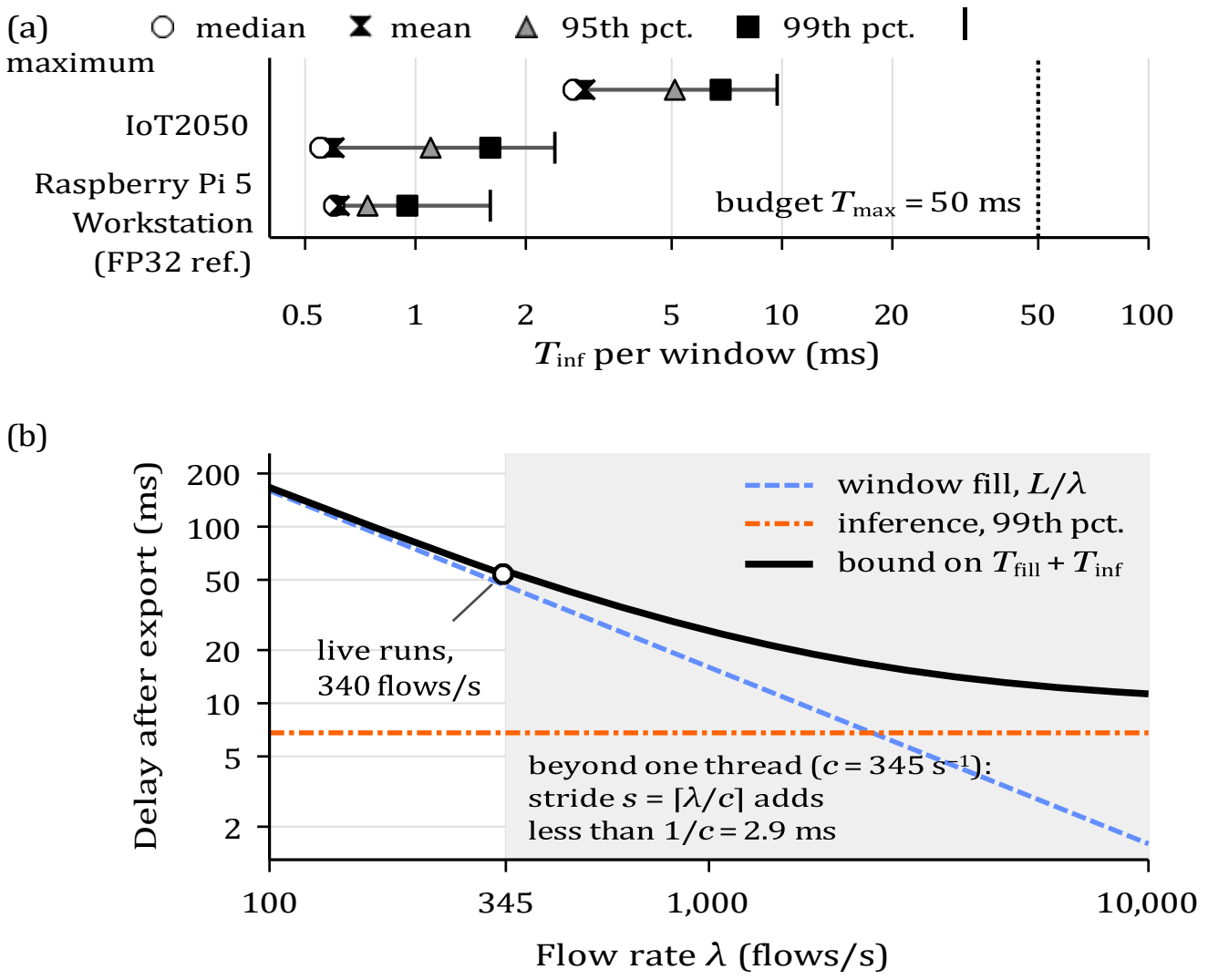

Fig. 6. (a) Per-window inference latency of the deployed INT8 model on the two gateways and of the FP32 reference on the workstation (Table X) against the budget. (b) Bounds on the window-fill and inference terms of the detection delay (1) on the IoT2050 as a function of the flow rate, computed from the measured 99th-percentile latency and single-thread throughput. Flow export and the cost of flow extraction are not included.

Raspberry Pi 5 0.6 ms and 1.6 ms. Both stay below the 50 ms budget, by factors of 7 and 31 (Fig. 6a). Peak resident memory is 31 MB and 28 MB. Energy is measured with an INA219 monitor across a 0.1 Ω shunt in the 5 V rail, sampled at 400 Hz, and is reported net of idle power as the mean over 1,000 consecutive inferences: 3.5 mJ and 1.3 mJ per inference. On the workstation the FP32 reference (PyTorch and scikit-learn) needs 0.58 to 0.79 ms for the GRU, 0.55 to 0.65 ms for the tree and 0.62 to 0.80 ms for the hybrid with parallel branches, depending on the dataset; that a tree of five comparisons costs as much as the GRU indicates that per-call framework overhead dominates at this model size.

*Capacity.* One inference thread on the IoT2050 sustains 345 windows per second, close to the flow rate of the live runs reported below (340 flows/s), and four threads reach 1,100. Scoring every $s$-th window divides the inference load by the stride $s$ and adds at most $(s-1)/\lambda$ to the detection delay. With $s = \lceil \lambda/c \rceil$ for a capacity of $c$ windows per second this addition is below $1/c$, one mean inference time. A stride of four, which changes F1 by at most 0.12 points on CIC-DDoS2019 and CICIoT23 (Section VI-H), therefore raises the flow rate that one thread can follow to about 1,380 flows/s.

## D. Detection of Modbus and MQTT Floods on the Testbed

Benign traffic is produced by `pymodbus` clients that poll a Modbus/TCP server with function code 3 every 100 ms and by MQTT publishers that send QoS-0 telemetry to a Mosquitto broker every second. The attacker then launches, in turn, a Modbus read flood, a Modbus write flood, a TCP SYN flood with `hping3`, an MQTT CONNECT flood, an MQTT PUBLISH flood and SlowITe. The hosts run on an emulated network (Mininet with Open vSwitch), and their traffic crosses a Cisco Catalyst 1000 aggregation switch, whose SPAN port mirrors it to the gateway (Fig. 1); the gateway itself is physical hardware in the loop. The traffic is captured with `tshark` and converted to flow records by a CICFlowMeter-compatible extractor on the gateway. Ground truth is assigned from the attacker address and the attack interval and is never a feature. Because the sensor is passive and works on flow features, the software endpoints limit the realism of the PLC-side attack surface, not of the sensing path.

TABLE XI
DETECTION PER ATTACK CLASS ON TESTBED TRAFFIC (IOT2050, DEPLOYED INT8 ARTEFACT OF SECTION VII-A, NO RETRAINING; VALUES IN %).

| **Attack class** | **Prec.** | **Rec.** | **F1** | **FPR** | **FNR** |
|---|---|---|---|---|---|
| Modbus read flood (FC 3) | 99.2 | 99.0 | 99.1 | 0.35 | 1.0 |
| Modbus write flood (FC 16) | 99.0 | 98.6 | 98.8 | 0.44 | 1.4 |
| TCP SYN flood, port 502 | 99.7 | 99.5 | 99.6 | 0.20 | 0.5 |
| MQTT CONNECT flood | 98.8 | 98.3 | 98.5 | 0.52 | 1.7 |
| MQTT PUBLISH flood | 99.4 | 99.1 | 99.2 | 0.30 | 0.9 |
| SlowITe (slow MQTT) | 95.8 | 94.4 | 95.1 | 1.20 | 5.6 |
| Macro average | 98.7 | 98.2 | 98.4 | 0.50 | 1.85 |

Table XI gives the detection rates per attack class. The five volumetric floods are detected with F1 between 98.5% and 99.6% at FPRs of 0.20% to 0.52%. SlowITe is the hardest case (F1 95.1%, FNR 5.6%): it imitates legitimate long-lived sessions. In ten further runs on the same set-up, with benign and attack traffic mixed, the sensor reached an accuracy of 98.9 ± 0.4% and an F1 of 99.0 ± 0.4% at an FPR of 0.71±0.12% and sustained 340±14 flows/s (mean ± standard deviation over runs). The processing time per flow on the gateway, from the arrival of a flow record at the detector to its fused score, was 3.1 ± 0.3 ms; it adds scaling and the window update to the 2.9 ms mean inference time of Table X. This is a per-flow processing time and part of $T_{inf}$ , not a detection delay: by (1) the delay also contains $T_{fill}$, up to 47 ms at this flow rate, and $T_{exp}$ (Section VII-E). These figures show that flow statistics learned from benchmark floods separate Modbus and MQTT floods from polling and telemetry traffic on this testbed. They carry the provenance caveat of Section VII-A, come without per-class flow counts or run-level intervals, and do not replace a repetition of the scenarios with the models of Section VI.

## E. Detection Delay

By (1) the delay to an alert has three parts. $T_{inf}$ is measured: at most 6.8 ms at the 99th percentile on the IoT2050, and 3.1±0.3 ms per flow, including scaling and the window update, in the live runs of Section VII-D. $T_{fill}$ is at most $L/\lambda$, that is 47 ms for $L$ = 16 at 340 flows/s, and it shrinks as a flood raises the flow rate; the symbolic branch can respond to the first exported attack record. Fig. 6b plots these bounds: at the rate of the live runs the fill term dominates, above about 2,400 flows/s the inference term does, and striding to keep one thread within its capacity adds less than 2.9 ms. $T_{exp}$ is set by the flow exporter, because a record exists only once its flow ends or a timeout expires. For volumetric floods, which create many short flows, it is small. For SlowITe, whose connections are held open, it is of the order of the exporter's timeout and dominates the delay. $T_{exp}$ was not instrumented in our runs, so no end-to-end time to detect is reported, and the 50 ms budget concerns $T_{\text{inf}}$ alone.

## VIII. Discussion and Limitations

*What the hybrid offers.* On this evidence the hybrid attaches a readable, stable rule to every decision of a small recurrent detector at no measurable cost in time or memory, with an operating point chosen under a stated budget. It lowers the FNR on CIC-DDoS2019 at the price of false alerts within that budget, and on the two informative benchmarks it does not beat simpler tree models on accuracy.

*Benchmarks.* None of the three benchmarks, as used here, provides interleaved traffic in capture order, so temporal modelling could not be tested, and the measured benefit of a window is that of pooling records of one class. Under the time-blocked scheme the attack windows of the CIC-DDoS2019 test partition come from two captures, and under the day-held-out scheme the comparison of the branches is confounded by one shuffled file. Edge-IIoTset is a consistency check, CICIoT23 measures single-record features, and one of those features is a probable artefact. The test partitions are attack-dominated, so the reading of precision that matters is that of Section VI-E.

*Selection.* The budget holds on validation data and only approximately on test data; rule (5) spends the whole budget for marginal gains in recall and can move weight to a weak branch when validation scores are saturated. A conservative order statistic [46] and a cost-aware tie-break are the natural remedies.

*Testbed.* The artefact on the gateway predates the evaluation protocol, the endpoints are software emulations on an emulated network, no production PLC or plant was involved, and the end-to-end detection delay was not measured. Latency, memory and energy are specific to the two Arm platforms and to the INT8 runtime measured.

*Scope.* The evaluation is closed-world [33]: attacks are drawn from the flooding families of the benchmarks and the six testbed scenarios, and robustness to evasion is not examined. The task is binary, and models are trained per dataset because the feature schemas of the benchmarks are mutually incompatible.

## IX. Conclusion

This paper presented a hybrid neural-symbolic DDoS detector for OT gateways, in which a compact GRU and a shallow decision tree are fused at score level under an explicit false-positive budget, and evaluated it with a leakage-controlled, dependence-aware protocol. Over ten seeds at a 1% budget the hybrid has an FNR of $0.08 \pm 0.01\%$ on CIC-DDoS2019, against 0.26% for the GRU and 0.17% for the tree, and it matches the tree on CICIoT23. With a budget of 0.1%, which we recommend, the FNR on CIC-DDoS2019 is 0.11% at about 313 false alerts per hour for 340 decisions per second. When a capture day with mostly unseen attack families is held out, the tree alone carries the decision at an FPR of 0.40% and an FNR of 3.13%. The audits show that the benchmarks as used do not allow temporal modelling to be tested, that Edge-IIoTset is saturated by duplicates, and that unconstrained trees, and on CICIoT23 tree ensembles, are more accurate than the hybrid. The case for the hybrid is therefore a short and stable rule for every decision together with a footprint measured on an industrial gateway.

Future work will repeat the testbed scenarios with the models evaluated here and with instrumented flow export, use physically commissioned PLC endpoints and OT captures with reliable timestamps, on which temporal modelling can be tested, and study differentiable neural-symbolic coupling, robustness to evasion and adaptive mitigation [6], [22].

## Acknowledgment

Generative AI was used to assist with restructuring and editing the text across all sections of the manuscript and with LaTeX formatting. The authors reviewed and approved the final content and take full responsibility for the manuscript.

## Code and Data

The notebook that produced all benchmark results, with its printed outputs, is available at https://github.com/MickeyKas/Neuro-SymbolicDDoS; it writes the split manifests, feature lists, rule sets and scores of every run.

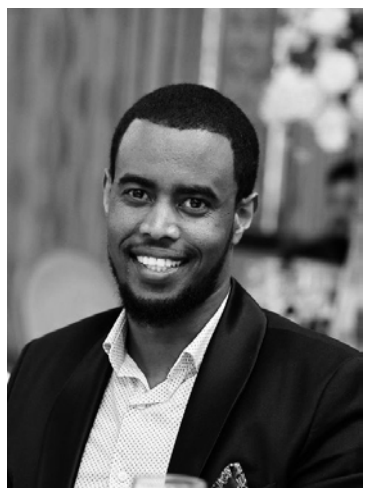

**Mikiyas Alemayehu** (Student Member, IEEE) received the B.Sc. (Hons.) degree in computer networking and cyber security with first-class honours and the M.Phil. degree in cyber security from London Metropolitan University, London, U.K. He is currently completing his Ph.D. in computer science, specialising in OT security, at Keele University, U.K. His previous experience includes research on an Innovate UK-funded project and technical training in networking. His research interests include industrial IoT, compliance, edge computing, digital forensics, and applied AI. Mr. Alemayehu is an Associate Fellow of the HEA (AFHEA) and holds several professional certifications, including ISC2 Certified in Cybersecurity (CC), Cisco CCNA, Cisco Security, and Cisco Ethical Hacking.

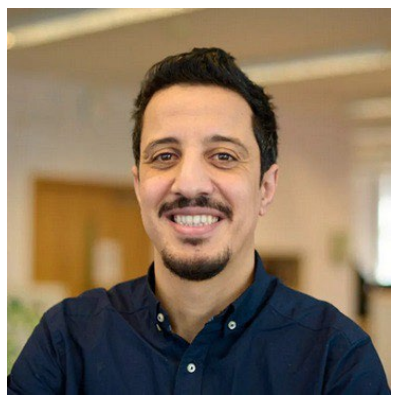

**Mohamed Chahine Ghanem** (Member, IEEE) is an Associate Professor and Cyber Security Research Lead at Keele University and a Visiting Associate Professor at the University of Liverpool. He is the Founder and CEO of ScrutinAIse, a start-up focused on agentic AI audit and accountability, and Principal Investigator of TOLLGATE-OT, an ARIA Safeguarded AI project developing a formally verified gateway between AI agents and operational technology. Before academia, he held senior positions in law enforcement and industry for over 15 years, including roles at Kroll and Nationwide, and remains a Senior Advisor on cyber resilience in banking. He holds an Engineering degree in Computer Systems from EMP, Algeria, an M.Sc. in Digital Forensics, and a Ph.D. in Cyber Security Engineering from City, University of London. His research interests include applied AI for cyber security, OT security, and digital forensics. He is a regular keynote speaker and Editor-in-Chief of the Journal of Cyber Security and Risk Auditing.

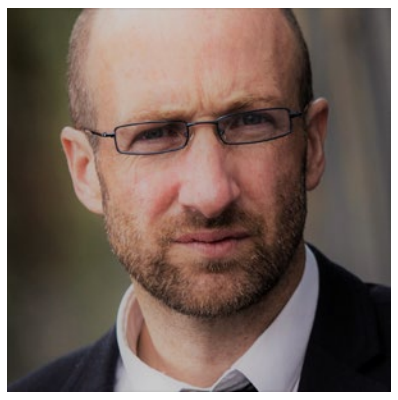

**Kevin Curran** (Senior Member, IEEE) is a Professor of Cyber Security at Ulster University and a recognised expert in cybersecurity and public engagement. He serves on the Advisory Group of the UK Cyber Security Council, the UK Border Vision Advisory Group, and the Northern Ireland Civil Service Cyber Leadership Board. He has contributed to numerous European and UK research projects and technology transfer schemes. His research has resulted in over 1,100 publications, with more than 12,500 citations and an h-index of 51 on Google Scholar. He is also a widely interviewed technology expert, with over 3,000 media interviews. Further details are available at https://kevincurran.org/interviews/.

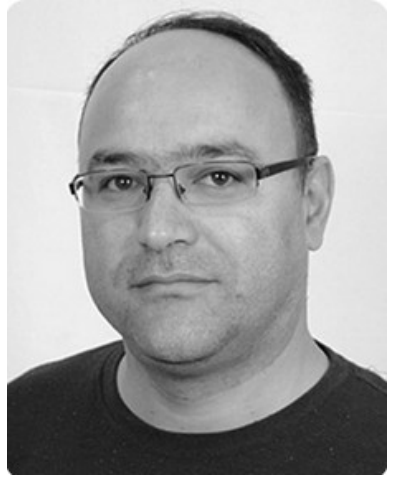

**Hamza Kheddar** (Senior Member, IEEE) received the master’s degree in telecommunications from ENSTTIC, the Ph.D. degree in telecommunications from USTHB University, and the Habilitation degree in 2021. He previously worked as a Core Network Expert at Huawei in the Middle East and North Africa region. He is currently an Associate Professor at the University of Medea and a Researcher with the LSEA Laboratory, Algeria. He has authored 30 research papers. His interests include speech processing, image classification, steganography, intrusion detection, digital twins, deep learning, 5G/6G security, biometrics, generative AI, and large language models. He has chaired telecommunications tracks at international conferences, including IC2M 2023 and ICTSS 2024, and serves as a reviewer for leading journals.

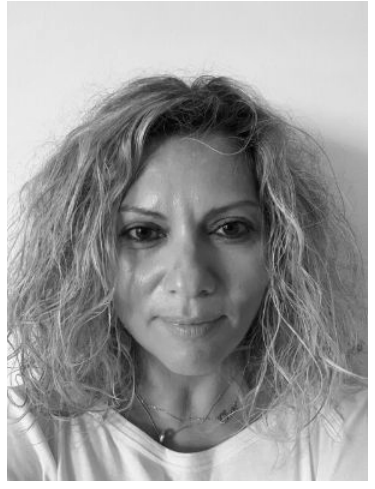

**Ioanna Dionysiou** (Senior Member, IEEE) received the B.Sc., M.Sc., and Ph.D. degrees in computer science from Washington State University, Pullman, WA, USA, in 1997, 2000, and 2006, respectively. She is a Professor of Computer Science at the University of Nicosia, Cyprus, where she serves as Associate Head of the Department of Computer Science and co-director of the Informatics Security Laboratory. Her doctoral research examined trust requirements for large-scale infrastructures, including the U.S. electric power grid, with support from NIST, NSF, and DOE. Her research interests include trust management, security and data protection in collaborative environments, active defences, security standards, and cyber warfare. She is a member of the Cyprus European Cyber Security Challenge Advisory Board and Vice-President of the ACM Cyprus Chapter.

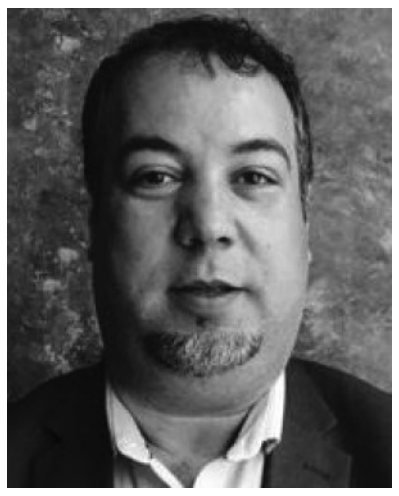

**Elhadj Benkhelifa** (Senior Member, IEEE) is a Full Professor of Computer Science and Director of the Smart Systems, AI and Cybersecurity Research Centre at the University of Staffordshire, U.K. He has published nearly 200 papers in international journals and conferences and chaired numerous international conferences. His research interests include biologically inspired algorithms, digital transformation, microservices, cloud/edge/mobile computing, AI, cybersecurity and resilience, software engineering, and data governance. He is an expert member of several UK and European Commission groups and committees, a member of the Alan Turing Institute’s Knowledge Graph Working Group, and Director of the Knowledge Graph Alliance. He is also a former Chair of the IEEE UK and Ireland Section’s Education Activities.

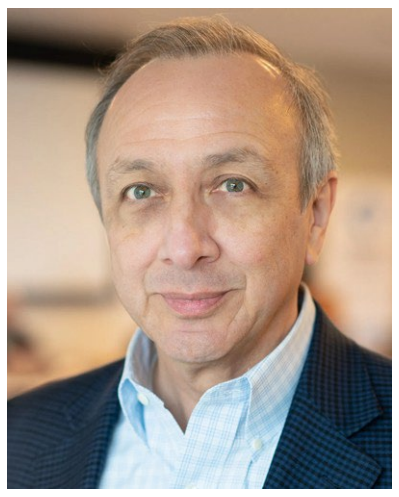

**J. J. Garcia-Luna-Aceves** (Fellow, IEEE) received the B.S. degree in electrical engineering from Universidad Iberoamericana, Mexico City, Mexico, in 1977, and the M.S. and Ph.D. degrees in electrical engineering from the University of Hawaii at Manoa, Honolulu, HI, USA, in 1980 and 1983, respectively. He is a Professor of Electrical and Computer Engineering and Canada Excellence Research Chair at the University of Toronto, Canada, and a Distinguished Professor Emeritus at the University of California, Santa Cruz, USA. He has held senior research and industry positions at the Palo Alto Research Center, SRI International, and Nokia. His research interests include network architectures, communication protocols, distributed algorithms, and performance analysis. He holds 70 U.S. patents and has published over 500 papers, which have received more than 43,000 citations. He has supervised over 40 Ph.D. theses and was the inaugural Chair of the ACM Special Interest Group on Multimedia.